\documentclass[sigconf]{acmart}

\usepackage{subfiles}

\usepackage{enumitem}
\usepackage{color, colortbl}
\usepackage{tikz}
\usepackage{caption}
\usepackage{subcaption}
\usepackage{multirow}
\usepackage{url}            
\usepackage{booktabs}       
\usepackage{amsfonts}       
\usepackage{nicefrac}       
\usepackage{amsthm}
\usepackage{makecell}
\usepackage{sc25repro}

\copyrightyear{2025}
\acmYear{2025}
\setcopyright{cc}
\setcctype{by}
\acmConference[SC Workshops '25]{Workshops of the International Conference for High Performance Computing, Networking, Storage and Analysis}{November
16--21, 2025}{St Louis, MO, USA}
\acmBooktitle{Workshops of the International Conference for High Performance
Computing, Networking, Storage and Analysis (SC Workshops '25), November
16--21, 2025, St Louis, MO, USA}
\acmDOI{10.1145/3731599.3767515}
\acmISBN{979-8-4007-1871-7/2025/11}

\begin{document}

\title{
    LLMTailor: A Layer-wise Tailoring Tool for Efficient Checkpointing of Large Language Models
}

\author{Minqiu Sun}
\affiliation{
  \institution{University of Delaware}
  \city{Newark}
  \country{United States}
}
\email{mqsun@udel.edu}

\author{Xin Huang}
\affiliation{
  \institution{RIKEN Center for Computational Science}
  \city{Kobe}
  \country{Japan}
}
\email{xin.huang@a.riken.jp}

\author{Luanzheng Guo}
\affiliation{
  \institution{Pacific Northwest National Laboratory}
  \city{Richland}
  \country{United States}
}
\email{lenny.guo@pnnl.gov}

\author{Nathan R. Tallent}
\affiliation{
  \institution{Pacific Northwest National Laboratory}
  \city{Richland}
  \country{United States}
}
\email{Nathan.Tallent@pnnl.gov}

\author{Kento Sato}
\affiliation{
  \institution{RIKEN Center for Computational Science}
  \city{Kobe}
  \country{Japan}
}
\email{kento.sato@riken.jp}

\author{Dong Dai}
\affiliation{
  \institution{University of Delaware}
  \city{Newark}
  \country{United States}
}
\email{dai@udel.edu}

\begin{abstract}

Checkpointing is essential for fault tolerance in training large language models (LLMs). However, existing methods, regardless of their I/O strategies, periodically store the entire model and optimizer states, incurring substantial storage overhead and resource contention. Recent studies reveal that updates across LLM layers are highly non-uniform. Across training steps, some layers may undergo more significant changes, while others remain relatively stable or even unchanged. This suggests that selectively checkpointing only layers with significant updates could reduce overhead without harming training. Implementing such selective strategies requires fine-grained control over both weights and optimizer states, which no current tool provides. To address this gap, we propose \texttt{LLMTailor}, a checkpoint-merging framework that filters and assembles layers from different checkpoints to form a composite checkpoint. Our evaluation indicates that LLMTailor can work with different selective checkpointing strategies and effectively reduce checkpoint size (e.g., 4.3 times smaller for Llama3.1-8B) and checkpoint time (e.g., 2.8 times faster for Qwen2.5-7B) while maintaining model quality.
\end{abstract}




\maketitle

\section{Introduction}

Originally developed in high-performance computing (HPC) to protect long-running scientific simulations, checkpointing has become the primary safeguard against failures in large-scale, high-cost LLM training. By periodically writing the complete model state, which includes weights, optimizer parameters, and other metadata, to persistent storage, the system can reload the most recent snapshot after a failure and resume training with minimal loss of progress~\cite{gandhi2025moetionefficientreliablesparse}.

Since current checkpointing uniformly saves all LLM layers, it introduces significant I/O overhead. As training scales up, the checkpoint-related overhead can account for 12\% of total training time and can rise to as much as 43\%~\cite{maengCPRUnderstandingImproving}. We believe this `saving the entire LLM states' approach might not be efficient, as many studies show that layers in LLMs are not updated at the same pace. 

Many recent works have highlighted the imbalances across layers of LLMs. Jawahar et al.~\cite{jawahar-etal-2019-bert} show that different layers of large language models encode distinct types of linguistic information. Phang et al.~\cite{phang-etal-2021-fine} observe that the lower and higher layers of fine-tuned RoBERTa and ALBERT models are dissimilar. Zhou et al.~\cite{temprature} propose the $PL\_Alpha\_Hill$ metric, demonstrating that different layers train at different speeds when using optimizers such as SGD and Adam. Collectively, these findings indicate that the existing checkpointing mechanisms, which indiscriminately save the complete model, are not optimal. It is then natural to explore the idea of keeping only part of the model during training. For example, checkpointing half of the layers at a time, and merging 2 checkpoints into a complete one. Clearly, doing so requires reconstructing a complete training state from multiple partial checkpoints, which in turn demands fine-grained manipulation of individual layers, including both weights and optimizer states. To our knowledge, no existing tool offers this capability. The closest one, MergeKit~\cite{mergekit}, merges the checkpoints only on weights and omits optimizers, making the recovery of training impossible. This is clearly not acceptable for checkpointing, as it is designed for training. 

In this study, we develop LLMTailor, a tool that merges both weights and optimizer shards to generate a fully resumable “Frankenstein” checkpoint, which can be used to continue training. With LLMTailor, we can investigate the feasibility and overhead of partial checkpointing during training.

To validate our design, we conduct extensive experiments to determine whether 1) partial checkpointing with LLMTailor reduces storage and checkpoint time, 2) the checkpoint made by LLMTailor will negatively impact the model quality; On Llama-3.1-8B, LLMTailor reduces total checkpoint size by 4.3 times; on Qwen-2.5-7B, we reduce the ratio of checkpoint time to end-to-end training time by 2.8 times. In another simple use case, we reduce the checkpoint size by half while maintaining model accuracy. The contributions of this study are threefold:

\begin{enumerate}
    \item We present LLMTailor, an effective tool that assembles a resumable “Frankenstein” checkpoint from parts of multiple checkpoints while maintaining the model performance. 
    \item We demonstrate that partial checkpointing with LLMTailor can substantially lower storage requirements and checkpoint time in LLM training.
    \item We show that LLMTailor incurs only a small amount of overhead when resuming training from a merged checkpoint.
\end{enumerate}

The rest of the paper is organized as follows. In \S\ref{sec:background} we discuss relevant backgrounds of LLMs. In \S\ref{sec:design} we describe the architecture of LLMTailor in detail. We present the extensive experimental results in \S\ref{sec:eval}. \S\ref{sec:related} discusses the closely related work. In \S\ref{sec:conclude} we present our conclusions and discuss future work.

\section{Background and Motivation}\label{sec:background}
Large language models (LLMs) have demonstrated remarkable power in various domains~\cite{naveed2025comprehensive,egersdoerfer2025stellar,egersdoerfer2025ioagent,egersdoerfer2024ion}. 
LLM training, like other neural networks, involves a forward pass, loss computation, a backward pass, and parameter update. Here, we review the key fundamentals that motivate our layer-wise checkpointing: (i) the structural components of LLMs, (ii) the organization of optimizers, and (iii) the distributed frameworks such as DeepSpeed used for scaling.

\subsection{Structure of LLMs}
\begin{figure}[ht!]
    \centering
    \includegraphics[width=\linewidth]{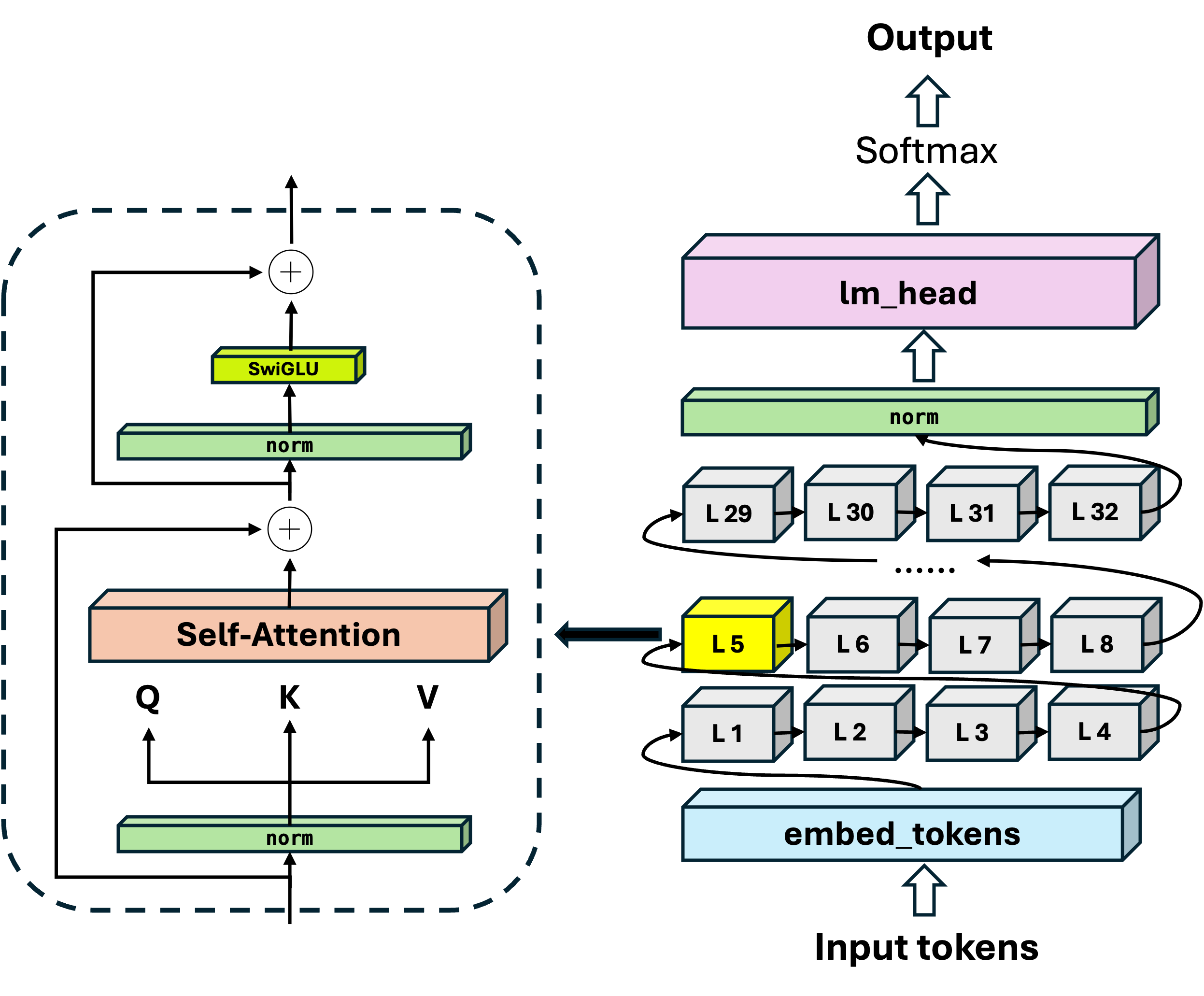}
    \caption{The layer-wise structure in a Llama3.1-8B model.}
    \label{fig:llama_workflow}
    \Description{The layer-wise structure in a Llama3.1-8B model.}
    \vspace{-1em}
\end{figure}

Figure~\ref{fig:llama_workflow} uses the Llama3-8B model as an example~\cite{llama3}: tokens are first embedded from a number to a vector, then propagated through 32 consecutive transformer layers. Then, after layer normalization to eliminate internal covariate shift, the hidden representations are projected back to the vocabulary space to produce logits in the \texttt{lm\_head} layer; finally, a softmax yields output probabilities. Note that not all models include a separate \texttt{lm\_head}; in smaller models, this layer is often weight-tied to \texttt{embed\_tokens} to reduce parameter count \cite{weight_tying}.

Each transformer block contains two layer-normalization sublayers that stabilize activation statistics. Between them, the self-attention module performs dynamic context aggregation, the defining operation of the Transformer architecture. The feed-forward network (FFN) then applies a non-linear transformation to each token representation: it expands the hidden dimension, activates it (e.g., with SwiGLU), and projects it back to the original size, thereby enriching the model’s ability to capture complex patterns. In the right panel of the figure, each box represents a trainable layer; collectively, these layers comprise the large language model.

\subsection{Structure of Optimizers}
The parameters are continuously updated to minimize the training loss, and this update process is applied by an optimizer, such as stochastic gradient descent (SGD) and Adam. Today, optimizers in training LLMs are often from the Adam family rather than plain SGD, due to their strong ability to accelerate and stabilize convergence during LLM training. As Equation~\ref{adam} shows, Adam maintains two running momentum estimates by continuously accumulating gradients into two auxiliary tensors $m$ and $v$ (the first- and second-order momentum). So, the optimizer has double the parameter size as the model weights due to the two momentum terms.

\begin{equation}
  \begin{aligned}
    W_i(t) &\gets W_i(t-1) - lr(t) \cdot \frac{m}{\sqrt{v}+\epsilon}.
  \end{aligned}
  \label{adam}
\end{equation}

\begin{figure}[ht!]
    \centering
    \includegraphics[width=0.8\linewidth]{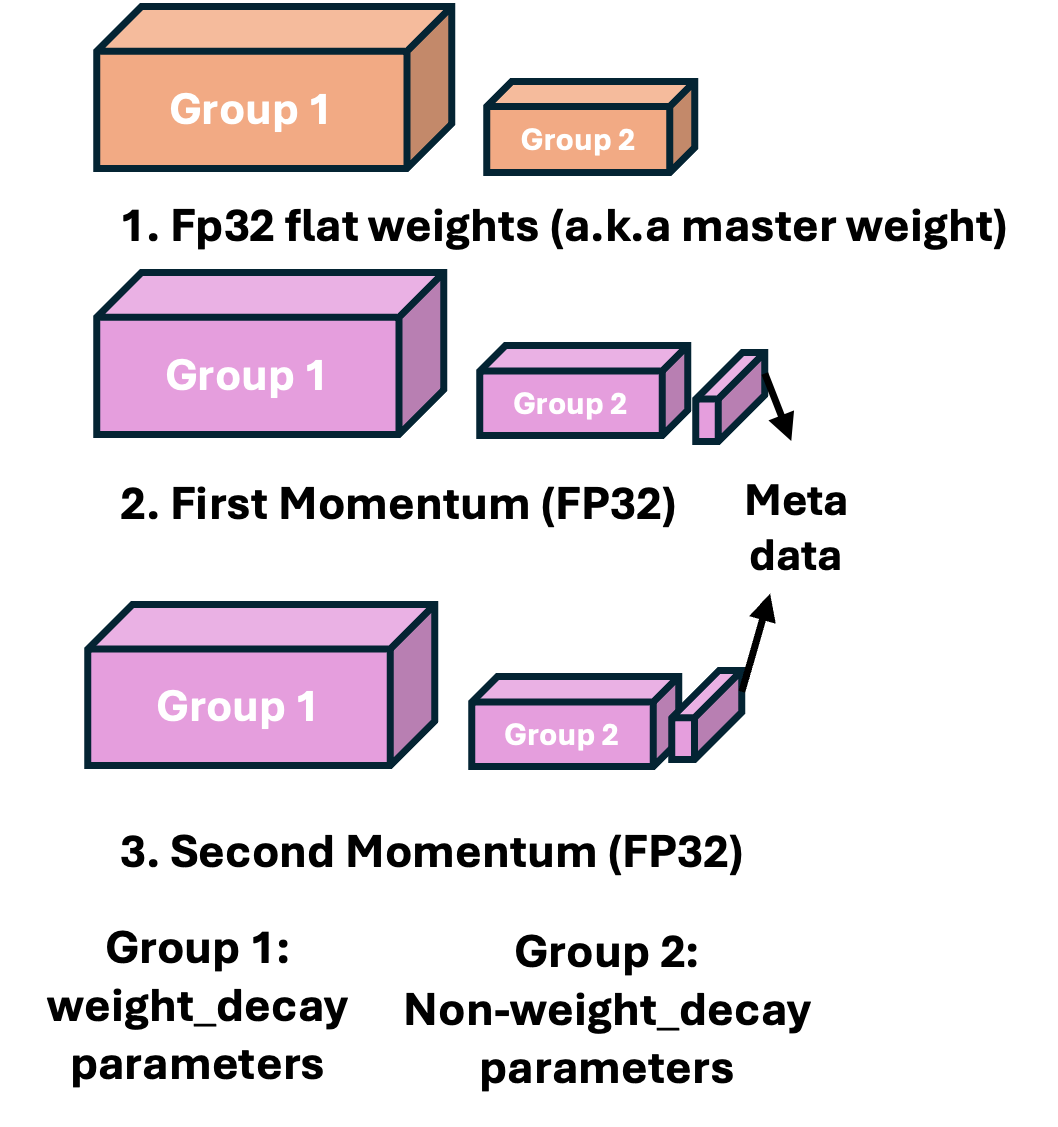}
    \caption{The AdamW optimizer used in a Llama3.1-8B model.}
    \label{fig:optimizer}
    \Description{The AdamW optimizer used in a Llama3.1-8B model.}
    \vspace{-1em}
\end{figure}

Figure~\ref{fig:optimizer} sketches the layout of the optimizer file in a checkpoint. During the update stage, all model parameters are flattened into two parameter groups, disregarding the model’s hierarchical or sub-layer structure, to improve computational efficiency. To preserve adequate optimization control, each group is assigned its own hyperparameters (e.g., learning rate, weight decay). Grouping at this coarse granularity rather than per tensor strikes a practical balance between efficiency and flexibility. In Figure~\ref{fig:optimizer}, the two groups are divided by different weight decay. This choice is because AdamW decouples weight decay from the loss gradient; it is standard to apply decay only to weights, while excluding biases and normalization parameters. Shrinking the latter can harm stability without offering meaningful regularization. Accordingly, one group contains all biases and normalization parameters (with zero weight decay), and the other contains the remaining weights (with nonzero weight decay).

Checkpointing safeguards training continuity by periodically snapshotting every component whose state evolves during optimization. The most obvious target is the model itself. In addition, the Adam optimizer relies on accumulated gradient moments to ensure precise and stable updates; omitting these optimizer states from the checkpoint can lead to large spikes or noisy updates when training resumes. In addition to the momentum terms, the file contains an FP32 master weight tensor of the same size. This duplication is required for mixed-precision training: forward and backward propagation run in FP16/BF16 to exploit tensor-core throughput and reduce memory-bandwidth demands, while FP32 master weights and FP32 momentum estimates are retained for numerical stability. Consequently, including some remaining configuration files, a single checkpoint must store at least 7 × the size of the FP16/BF16 model itself.

\subsection{Distributed LLM training under Deepspeed}
Since LLMs have now grown to unprecedented scales, this footprint makes distributed frameworks such as DeepSpeed’s ZeRO vital to LLM training. It can shard optimizer states, gradients, and parameters across data-parallel processes, dramatically reducing the per-GPU memory footprint \cite{zero3,deepspeed}. For example, in the ZeRO3 stage, each rank holds only a shard of the model parameters and optimizer states. During forward/backward passes, ZeRO-3 all-gathers the needed parameter shards just-in-time for a layer, computes them, and then re-shards them. In distributed training checkpoint systems, optimizer states are saved as shards: each GPU writes only its own shard to reduce overhead. In contrast, to ensure that it can be used for reasoning at any time, the model weights are typically stored as a single consolidated file. During recovery, every optimizer shard is loaded onto its corresponding GPU, ensuring that the training state is restored correctly.

\section{Brief Introduction about Mergekit}

Mergekit~\cite{mergekit} is a lightweight command-line toolkit that lets users compose new language-model checkpoints from two or more existing ones with only a short YAML recipe. Internally, it loads each source model, matches tensors by name, and applies a user-selected merge rule. To use it, users first select a merge method, such as \emph{linear} blending, \emph{SLERP}, \emph{passthrough} copy, or \emph{LoRA‐fusion}. Next, users need to write a recipe YAML that lists source models and specifies layer-selection rules. And then run the CLI to emit a new model assembled according to these rules. After that, the resulting Frankenstein model can be loaded directly by standard PyTorch or Hugging Face runtimes. In practice, users most often apply it to tasks like model capability fusion and domain adaptation.

Although mergekit supports several merging strategies, layer-wise checkpoint merging relies on the passthrough method, which splits and recombines source models by layer. Because MergeKit works only with weight files and therefore avoids both back propagation and retraining, it cannot support full checkpoint merging for the following reasons:
 \begin{enumerate}
     \item Optimizer states are ignored. Mergekit merges only model weights, omitting optimizer files that are essential for resuming training.
     \item Auxiliary layers are excluded. It manipulates transformer layers only, leaving out large auxiliary layers such as token embeddings and the prediction head.
     \item Configuration files are not handled. Mergekit does not provide support for merging or editing checkpoint configuration files.
 \end{enumerate}
 
Given the popularity and practicality of MergeKit, we want to adopt its YAML-driven interface and extend its functionality to handle full checkpoints, including optimizer states, auxiliary layers, and configuration metadata. 

\section{Design and Implementation}\label{sec:design}

In this section, we introduce LLMTailor, which constructs resumable training checkpoints by composing model layers (and associated optimizer states) from multiple checkpoints without changing the way mergekit is used. The key design insights behind LLMTailor are detailed below.

\subsection{Construct Separable Optimizers in Checkpoint}

\begin{figure}[ht!]
\Description{}
    \centering

    \includegraphics[width=\linewidth]{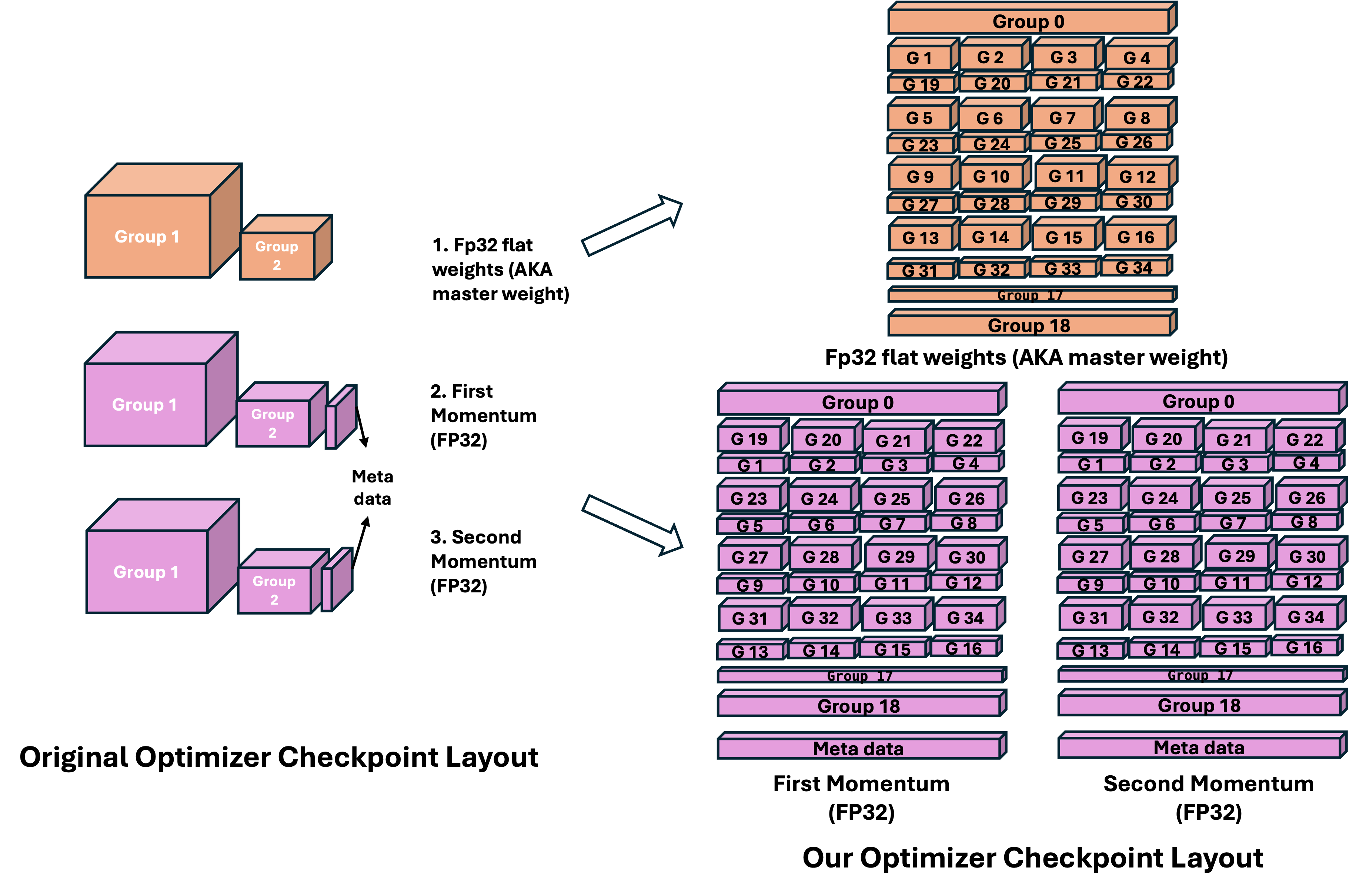}
    \caption{Reconstruct the parameter groups in the optimizer before training.}
    \label{fig:optimizer_rebuild}
    \vspace{-1em}
\end{figure}

Because MergeKit cannot merge optimizer states from different checkpoints, our LLMTailor extends it to enable this functionality. The main challenge arises from the optimizer’s intrinsic structure: unlike model weights, which follow a layer-wise hierarchy, optimizer files store flattened tensors that are difficult to split or merge. The only natural partition point in these files is the parameter group. We therefore reconstruct the parameter groups to mirror the model’s layer-wise organization while preserving the original weight-decay settings. Specifically, each transformer layer is divided into two groups: one containing tensors exempt from weight decay and the other containing the remaining tensors. Since auxiliary layers contain exclusively either weight-decay or non-weight-decay parameters, they are assigned to a single parameter group. Since the ordering of such parameter groups is consistent across different LLMs, knowing only the total number of transformer layers and whether weight tying is applied by reading the configuration file is sufficient to determine the parameter group index of each layer in the optimizer file. As a result, the number of parameter groups increases from two to $2L + x$, where $L$ is the number of transformer layers, $x$ is the number of auxiliary layers. Figure \ref{fig:optimizer_rebuild} illustrates the transformation of a 16-layer, 2-group model into a 35-group model.

The checkpoint’s structure must match the model in GPU memory during training or before failure; otherwise, the checkpoint is unusable. As a result, we perform this regrouping before training begins. During training, the master weights are automatically organized into the same 35 groups, ensuring that the optimizer file stored in each checkpoint has a uniform, group-aligned structure. Next, we merge and assemble a new checkpoint by tracking the indices of the two parameter groups associated with each transformer layer.

This redesign renders the optimizer files separable, while preserving the original weight decay configuration. Because neither parameters nor hyperparameters are altered, the training procedure and final results remain unchanged; the only additional cost is a small amount of computational overhead. This helps solidify the basis for tailoring checkpoints.



\subsection{Merge Optimizers}

Supporting arbitrary assembly of layers from multiple checkpoints, especially in the optimizer files, is a key objective. Since MergeKit already handles weight merging, we focus here on our implementation of the optimizer file.

LLMTailor first parses a YAML specification that lists the base model, the source layers with their corresponding checkpoints, and the target positions of those layers in the new model. The fixed parameter-group structure lets us locate each layer and its associated groups. Figure \ref{fig:optimizer_rebuild} illustrates the default ordering of parameter groups: the first group stores the normalization layer, followed by the non-weight-decay segments of the transformer layers, then the embedding layer and the optional \texttt{lm\_head}, and finally the weight-decay segments of each transformer layer. When a transformer layer is selected, LLMTailor automatically indexes and copies all its parameter groups, inserting them into the user-defined position in the new model.

Unlike the unified model-weight file, each parameter group is uniformly sharded across GPUs, and the corresponding checkpoints are stored in separate files because of distributed training. The largest overhead in LLMTailor is the I/O overhead of loading up to $N\times(L+3)$ optimizer files and writing up to $N$ files, where $N$ represents the total number of GPUs. To ensure the correctness of the resumed checkpoint, we keep the order of loading and writing. In addition, we leverage multiprocessing to accelerate this process. We employ Python’s ProcessPoolExecutor to parallelize shard loading across multiple CPU cores, enabling concurrent decompression and deserialization of large ZeRO-3 optimizer files, which significantly reduces overall I/O latency.

\subsection{Split Auxiliary Layers}
Since mergekit focuses solely on model weight merging, it retains the base model’s \texttt{norm}, \texttt{embed\_token}, and \texttt{lm\_head} for vocabulary compatibility. However, the latter two contain the entire vocabulary and thus occupy far more space than individual transformer blocks. We therefore adjust the merge plan to split and merge these auxiliary modules explicitly. In \textsc{LLMTailor}, the only extra requirement is that users list these three modules in the YAML recipe.

\subsection{Manipulate Configuration Files}
Metadata and configuration files record user-configured arguments, training state history, the current training step, and the current learning rate. Thus, these files must be copied from the most recent checkpoint to assemble a new "Frankenstein" checkpoint while preserving training continuity and effectiveness. \textsc{LLMTailor} also provides an autonomous copy of these configuration files in the latest checkpoint that we create. 

\section{Experiments}\label{sec:eval}
In this section, we describe the experimental setup, models, datasets, baselines, and give 2 use case for our LLMTailor. 

\begin{table*}[ht!]
    \centering
    \begin{subtable}[b]{0.49\textwidth}
      \centering
      \small
      \begin{tabular}{@{} l c c @{}}
        \toprule
        \textbf{Model} & \textbf{Final train loss} & \textbf{Final eval loss} \\
        \midrule
        Qwen2.5-7B (After SFT) & 1.58 & 1.60 \\
        Parity merge (start from 400)    & 1.58 & 1.60 \\
        \bottomrule
      \end{tabular}
      \caption{Qwen2.5-7B in a SFT task.}
      \label{tab:final-loss-qwen}
    \end{subtable}
    \hfill
    \begin{subtable}[b]{0.49\textwidth}
      \centering
      \small
      \begin{tabular}{@{} l c c @{}}
        \toprule
        \textbf{Model} & \textbf{Final train loss} & \textbf{Final eval loss} \\
        \midrule
        Llama3.1-8B (After CPT) & 1.58  & 1.58 \\
        Filtered Layers (start from 1000)    & 1.58  & 1.58 \\
        \bottomrule
      \end{tabular}
      \caption{Llama3.1-8B in a CPT task.}
      \label{tab:final-loss-llama}
    \end{subtable}
    \caption{Comparison of training loss between original checkpoint and new checkpoint created by parity.}
    \label{tab:result1}
\end{table*}

\begin{table*}[ht!]
  \centering
  \small
  \begin{tabular}{@{} r c c c c c c @{}}
    \toprule
    Task    & Model & MMLU & MMLU\_med & MedMCQA & MedQA & PubMedQA \\
    \midrule
    \multirow{2}{*}{SFT}
    & Qwen2.5-7B      & 73.14 & 89.00          & 60.75 & 64.02          & 75.20 \\
    & parity-400      & 72.89 & 87.00          & 60.58 & \textbf{64.10} & \textbf{76.20} \\
    \midrule
    \multirow{2}{*}{CPT}
    & Llama3.1-8B & 60.00          & 75.00   & 53.10          & 55.15       & 77.20 \\
    & parity-1000 & \textbf{60.03} & 72.00   & \textbf{53.12} & 54.36       & 76.60 \\
    \bottomrule
  \end{tabular}
  \caption{Zero‐shot Benchmark Evaluation Results of resumed training in use case 1. (Question-Answer, higher is better)}
  \label{tab:case1_bench}
  \vspace{-1em}
\end{table*}

\subsection{Experimental Setting}
We conducted all real-world experiments on an 8-GPU cluster featuring NVIDIA A100 80GB GPUs and two AMD EPYC 7713 CPUs. The file system that we use is a Lustre file system mounted via an InfiniBand network. The used software versions are CUDA 12.8, DeepSpeed v0.17.2, and PyTorch 2.7.1. To address GPU memory constraints and train larger models, we employed DeepSpeed ZERO Stage-3 optimization, with the default optimizer being AdamW~\cite{adamw}. 

We evaluate LLMTailor using popular open-source LLMs, namely Llama-3.2-1B, Llama-3.1-8B~\cite{llama3}, and Qwen-2.5-7B~\cite{qwen2025qwen25technicalreport}. All experiments employ a sequence length of 2048. To quantify the reduction in checkpoint overhead in post-training, we test LLMTailor on two representative tasks: continual pre-training (CPT)~\cite{cpt} and supervised fine-tuning (SFT)~\cite{sft}. Because post-training typically adapts a general-purpose LLM to specialized knowledge or capabilities, we select two medical datasets: (i) PubMed-Summarization, a plain-text corpus for CPT~\cite{pub}, and (ii) MedQA~\cite{medqa}, a structured question-answering dataset for SFT. Each task is trained for one epoch. For PubMed-Summarization, we use a micro-batch size of 4 with 2 gradient-accumulation steps; for MedQA, we use a micro-batch size of 2 with 2 gradient-accumulation steps. 

To assess model quality after recovery, we evaluate it on five benchmarks spanning three categories: medical expertise, general knowledge, and reasoning. Our goal is not to improve absolute scores but to verify that LLMTailor does not degrade performance in the post-training context. Consequently, we do not perform dataset deduplication before evaluation. Since no prior work has explored partial checkpointing, and partial checkpointing mechanisms can also be combined with prior work on I/O optimization and in-memory techniques, as the approaches are not mutually exclusive, we establish our baseline using the default checkpointing mechanism provided by the \textit{transformers} library. Specifically, we adopt fixed checkpoint intervals of every 50 steps for the SFT task and every 100 steps for the CPT task, and then compare the time and storage overhead of our method against this baseline. 

\subsection{Use Case 1: Merge Checkpoints by Parity}

The first use case that we decide to test is to merge the odd layers and the \texttt{embed\_token} layer from the previous checkpoint, and the even layers and the \texttt{lm\_head} layer from the current checkpoint. Considering the layer-wise structure of LLM is sequential, in order to avoid the split of the front and back layers of the checkpoint, we use this method to reconstruct a new one for resuming. Checkpointing only half of the complete checkpoint each time can reduce the storage overhead by almost half. Table \ref{tab:case1} lists the exact size of each checkpoint.

\begin{table}[ht]
  \centering
  \small
  \begin{tabular}{@{} r c c c c @{}}
    \toprule
    Model      & Type & Total CKPT size (G) & \makecell{The proportion of\\checkpoint time(\%)}\\
    \midrule
    \multirow{2}{*}{Llama3.1-8B}   & Total  & 1799.52 & 4.99\\
                                   & \textbf{Parity}   & \textbf{899.76}  & \textbf{3.03}  \\
    \multirow{2}{*}{Qwen2.5-7B}    & Total  & 1811.52 & 20.63\\
                                   & \textbf{Parity}   & \textbf{905.76}  & \textbf{12.76} \\
    \bottomrule
  \end{tabular}
  \caption{Comparison between complete checkpoint and partial checkpoint in parity checkpoint}
  \label{tab:case1}
  \vspace{-1em}
\end{table}

In Table \ref{tab:case1}, the proportion of checkpoint time means the ratio of checkpointing time and the end-to-end training time. These results suggest that in this use case, we can reduce the storage overhead by approximately 50\% and checkpointing time overhead by 40\%. 

We then resume training using the new checkpoints generated by our LLMTailor framework. Table~\ref{tab:result1} compares the resumed training process with that of the original model. The results show that both the training loss and the evaluation loss at the final step match those of the original training trajectory. This demonstrates the correctness of the checkpoint merging in this use case.

\begin{table*}[ht!]
    \centering
    \begin{subtable}[b]{0.49\textwidth}
      \centering
      \small
      \begin{tabular}{@{} l c c @{}}
        \toprule
        \textbf{Model} & \textbf{Final train loss} & \textbf{Final eval loss} \\
        \midrule
        Qwen2.5-7B (After SFT) & 1.58 & 1.60 \\
        Filtered Layers (start from 400)  & 1.60 & 1.62 \\
        \bottomrule
      \end{tabular}
      \caption{Qwen2.5-7B in an SFT task.}
      \label{tab:final-loss-qwen1}
    \end{subtable}
    \hfill
    \begin{subtable}[b]{0.49\textwidth}
      \centering
      \small
      \begin{tabular}{@{} l c c @{}}
        \toprule
        \textbf{Model} & \textbf{Final train loss} & \textbf{Final eval loss} \\
        \midrule
        Llama3.1-8B (After CPT) & 1.58 & 1.58 \\
        Filtered Layers (start from 1000)    & 1.59 & 1.59 \\
        \bottomrule
      \end{tabular}
      \caption{Llama3.1-8B in a CPT task.}
      \label{tab:final-loss-llama1}
    \end{subtable}
    \caption{Comparison of training loss between original checkpoint and new checkpoint created by the impact of layers.}
    \label{tab:result2}
    \vspace{-1em}
\end{table*}

\begin{table*}[ht!]
  \centering
  \begin{tabular}{@{} r c c c c c c @{}}
    \toprule
    Task    & Model & MMLU & MMLU\_med & MedMCQA & MedQA & PubMedQA \\
    \midrule
    \multirow{2}{*}{SFT}
    & Qwen2.5-7B      & 73.14 & 89.00  & 60.75 & 64.02  & 75.20 \\
    & filter-400      & 71.64 & 84.00  & 59.50 & 62.06  & \textbf{75.60} \\
    \midrule
    \multirow{2}{*}{CPT}
    & Llama3.1-8B & 60.00          & 75.00          & 53.10          & 55.15          & 77.20 \\
    & filter-1000 & \textbf{62.06} & \textbf{77.00} & \textbf{53.45} & 54.91          & \textbf{78.00} \\
    \bottomrule
  \end{tabular}
  \caption{Zero‐shot Benchmark Evaluation Results of resumed training in Use case2. (Question-Answer, higher is better)}
  \label{tab:case2_bench}
  \vspace{-1em}
\end{table*}
To assess the quality of the model after recovery and continued training, we evaluate the final, fully trained models on five benchmarks spanning three domains: medical expertise, general knowledge, and reasoning. Table~\ref{tab:case1_bench} reports these results, where higher scores indicate better performance. The top results for each benchmark are highlighted. Here, the absolute score might not be the most important thing. Instead, it is crucial to verify that the Frankenstein model does not significantly degrade performance compared to the original model, which never experiences any failures or recoveries. 
Results show that our method can preserve model quality after recovering from the merged checkpoints. Therefore, it offers the possibility that future checkpointing systems could incorporate more precise mechanisms that allow for a slight trade-off in model performance to better balance overall model quality with the reduction in overhead.

\subsection{Use Case 2: Merge Checkpoints by Filtering}
Previous work shows that the first several layers and the last two layers of the model have a greater impact on model reasoning~\cite{gromov2025unreasonableineffectivenessdeeperlayers}. Based on that, we decide to checkpoint by filtering only the first and the last 2 layers each time, and checkpoint half of the other layers less often $ every 5\times original\_interval$-to- to reduce more overhead. Table \ref{tab:case2} lists the exact size of each checkpoint. 

\begin{table}[ht!]
  \centering
  \small
  \begin{tabular}{@{} r c c c c @{}}
    \toprule
    Model      & Type & Total CKPT size (G) & \makecell{The proportion of\\checkpoint time(\%)}\\
    \midrule
    \multirow{2}{*}{Llama3.1-8B}   & Total    & 1799.52 & 4.99\\
                                   & \textbf{Filtered}   & \textbf{420}     & \textbf{1.66} \\
    \multirow{2}{*}{Qwen2.5-7B}    & Total    & 1811.52 & 20.63\\
                                   & \textbf{Filtered}   & \textbf{434.56}  & \textbf{7.26} \\
    \bottomrule
  \end{tabular}
  \caption{Comparison between complete checkpoint and partial checkpoint in filtered checkpoint}
  \label{tab:case2}
\end{table}

Here, we also set the default checkpoint mechanism as a baseline and measure the proportion of time spent on checkpoints. In this scenario, we can reduce the checkpoint time ratio to up to 2.8 × for the Qwen2.5-7B model, and 4.3 × storage overhead for the Llama3.1-8B model. We compare the size of the checkpoint with the default checkpoint method. Then, we use the new checkpoint created by our LLMTailor to resume training. Table~\ref{tab:result2} shows the results of their continued training compared to the original model.

We then resume training using the new checkpoints generated by our LLMTailor framework. Table~\ref{tab:result2} compares the resumed training process with that of the original model. Since lower loss indicates better performance, the results show that both the training loss and the evaluation loss at the final step decrease slightly. This demonstrates that performance could degrade if we focus solely on reducing overhead at the expense of preserving the original functionality of checkpoints.

In the benchmark evaluation, Table~\ref{tab:case2_bench} presents the results, where higher scores indicate better performance, and the top score for each benchmark is highlighted. In the SFT task, the newly created Qwen2.5 checkpoint performs noticeably worse than the default checkpoint, whereas in the CPT task, the composite Llama3 model noticeably outperforms the default model. This suggests that the inherent robustness of LLMs can, to some extent, support our concept of partial checkpointing. These relatively strong results from the rule-based partial checkpointing mechanism suggest that future systems employing more dynamic strategies in deciding which components to checkpoint and when are likely to achieve even better performance and greater robustness.

\subsection{Checkpoint Overhead}
We evaluate LLMTailor for splitting and merging different LLMs on both CPT and SFT tasks. The checkpoints we merge are partial. For the baseline, the reported time corresponds only to resuming a checkpoint, whereas for the other checkpoints, the reported time also includes the additional overhead of running LLMTailor. Therefore, in Table~\ref{tab:overhead}, we present the measured merging overhead from different perspectives.

\begin{table}[ht!]
  \centering
  \footnotesize
  \begin{tabular}{@{} r c c c c @{}}
    \toprule
    Model Name  & Checkpoint Size (G) & Total layers & CKPTs included & Time (s) \\
    \midrule
    \multirow{4}{*}{Llama3-1B}   & \multirow{4}{*}{17.29}  & \multirow{4}{*}{18}  & Baseline: 1 & 0.80 \\
      &  &  & 2 &   117  \\
      &  &  & parity (2) & 233.6 \\
      &  &  & 8 &  60.4 \\
      &  &  & 18 &  62.5  \\
    \midrule
    \multirow{4}{*}{Llama3-8B}   & \multirow{4}{*}{112.47}  & \multirow{4}{*}{35} & Baseline: 1 & 16.8\\
      &  &  & 2 &  332.4  \\
      &  &  & parity (2) & 1027.5 \\
      &  &  & 8 &  279.2  \\
      &  &  & 35 & 264.3   \\
    \bottomrule
  \end{tabular}
  \caption{Loading time for different checkpoints}
  \label{tab:overhead}
  \vspace{-1em}
\end{table}

Unless otherwise specified, checkpoints are loaded in a straightforward manner; for example, layers(1\text{–}16) are loaded from checkpoint-100, layers (17\text{–}32) from checkpoint-200, and so on. When we refer to “parity,” however, checkpoints are loaded multiple times in an interleaved fashion: first, layer(1) from checkpoint-100 and layer(2) from checkpoint-200; then layer(3) from checkpoint-100 and layer(4) from checkpoint-200, and so forth. Even with only two checkpoints, this process still requires loading and discarding them N times, where N is the total number of layers. This approach incurs substantial overhead because the optimizer state can only be accessed after the checkpoint is fully loaded, with no possibility of lazy loading, as in the case of model weights. From this, we observe that the time overhead in LLMTailor is determined by: i) the loaded checkpoint size; ii) the number of loaded checkpoints; iii) the method we load the layers; iv) the number of total layers. Compared with the total training time, which can span several hours or even days, this overhead is relatively small and thus acceptable. We also observe that, although checkpoints must be loaded from N different files, each checkpoint contains only a single layer. As a result, the loading process is relatively fast. This observation suggests that the overhead of LLMTailor could be significantly reduced once a layer-wise checkpointing system is adopted.

\section{Related Work}\label{sec:related}
\subsection{Checkpointing in Deep Learning.}
Checkpoint techniques have been widely explored and utilized in many deep learning frameworks, such as PyTorch~\cite{paszkePyTorchImperativeStyle2019}, TensorFlow~\cite{abadiTensorFlowSystemLargescale2016}. However, default checkpoint systems will introduce significant overhead and stalls during the training stage, especially for large models. To address this challenge, many works have proposed different optimizations. Some works focus on optimizing the checkpoint system, such as CheckFreq~\cite{checkfreq}, which dynamically adjusts the checkpointing frequency to reduce I/O; Check-N-Run~\cite{check-n-run} compresses checkpoints with lossy schemes to reduce both I/O and required storage;~\cite{maengCPRUnderstandingImproving,partialrecover} recovers only checkpoints on GPUs that failed. Gemini~\cite{wangGEMINIFastFailure2023} introduces in-memory checkpoint protection to avoid stalls and reduce recovery time by using high-bandwidth CPU memory, while Just-in-Time checkpointing~\cite{guptaJustInTimeCheckpointingLow2024} utilizes state redundancy in data parallel replicas of large deep learning jobs for efficient run-time checkpointing. In addition, Datastate-LLM~\cite{Datastate-LLM} uses a lazy asynchronous multi-level approach, optimizing the pipeline of the checkpoint system to reduce overhead.

However, unlike all these methods, our approach focuses on identifying the non-uniform weight updates in LLM training and do layer-wise checkpointing to reduce the overhead, so that we can further reduce checkpoint size while utilizing current optimizations.



\subsection{Model Merging}
Researchers also disassemble and merge models. For instance, model merging is widely studied as an effective training-free method to combine the capabilities of fine-tuned large language models~\cite{wortsmanModelSoupsAveraging2022}. Yadav et al.~\cite{yadavTIESMergingResolvingInterference2023} further help solve conflicts between different task vectors through norm-based sparsification and consensus on the signs of weights. Yu et al.~\cite{yuLanguageModelsAre2024} showed that LLMs are highly robust to sparsifying task vectors and proposed DARE for merging LLMs with random sparsification. Recently, Online Merging Optimizers~\cite{luOnlineMergingOptimizers2024} online merge the gradients of the policy model with the delta parameters of the SFT model. Our experiments show that both the model and the optimizer in checkpoints can also benefit from model merging.

\section{Conclusion}\label{sec:conclude}

In this paper, we introduce LLMTailor, a tool that enables effective and lightweight layer-wise checkpoints split-and-merge for large models. By performing layer-wise checkpointing, our use cases show the potential of reducing total checkpoint size and time by at least 60\% compared with existing solutions, while preserving final model quality. Although our prototype can only manipulate local checkpoints, the underlying design can be applied to other checkpointing frameworks, and the implementation of this is part of our future work.

\section*{Acknowledgements}
We sincerely thank the anonymous reviewers for their valuable feedback. This work was supported in part by the National Science Foundation (NSF) under grants CNS-2008265 and CCF-2412345. This effort was also supported in part by the U.S.\@ Department of Energy (DOE) through the Office of Advanced Scientific Computing Research's ``Orchestration for Distributed \& Data-Intensive Scientific Exploration'' and
the ``Decentralized data mesh for autonomous materials synthesis'' AT SCALE LDRD at Pacific Northwest National Laboratory. PNNL is operated by Battelle for the DOE under Contract DE-AC05-76RL01830. 

\bibliographystyle{ACM-Reference-Format}
\bibliography{references}

\appendix

\section{Overview of Contributions and Artifacts}

\subsection{Paper's Main Contributions}

\begin{description}
\item[$C_1$] We present LLMTailor, an effective tool that assembles a resumable checkpoint from parts of multiple checkpoints while maintaining the model performance. 
\item[$C_2$] We demonstrate that partial checkpointing with LLMTailor can substantially lower storage requirements and checkpoint time in LLM training.
\item[$C_3$] We show that LLMTailor incurs only a small amount of overhead when resuming training from a merged checkpoint.
\end{description}

\subsection{Computational Artifacts}

\begin{description}
\item[$A_1$] https://doi.org/10.5281/zenodo.16909083
\end{description}

\begin{center}
\begin{tabular}{rll}
\toprule
Artifact ID  &  Contributions &  Related \\
             &  Supported     &  Paper Elements \\
\midrule
$A_1$   &  $C_1, C_2, C_3$ & Tables 1-7 \\
        &        & Figure 3\\
\bottomrule
\end{tabular}
\end{center}

\section{Artifact Identification}

\newartifact

\artrel

The artifact provides the implementation of the methods and ideas presented in the paper. Our LLMTailor, included as the artifact, supports fine-grained manipulation of model and optimizer states across checkpoints, selective merging part of the checkpoints, and reconstruction of training state. It is also the key to realizing the proposed checkpoint strategies as it enables partial layer saving.

\artexp

\begin{enumerate}
    \item \textbf{Flexibility in Checkpoint Composition} By using LLMTailor, we are able to assemble “Frankenstein” checkpoints. It selects and mixes layers from different checkpoints and shows both functional correctness (the model runs and trains as expected) and practical utility (supporting adaptive failure recovery or analysis).
    \item \textbf{Reduced Checkpointing Overhead} The experiments will show that LLMTailor enables selective, layer-level checkpointing which can reduce the size and time of saved checkpoints. Compared to traditional full-model checkpoints, the system achieves significant I/O savings and reduced storage usage without sacrificing recoverability.
    \item  \textbf{Faithful Model Recovery and Continuity} By merging layers and optimizer states from multiple checkpoints, use case 1 (merge by parity) demonstrates that training can be faithfully resumed from partial checkpoints. The outcome to expect is a recovery trajectory that closely matches (or even exactly overlays) the trajectory from full checkpointing baselines. While in use case 2 (filter layers), the recovered training will have a bias with the original checkpoint.
\end{enumerate}

\arttime

The expected computational time of this artifact on GPU is 60~min.

\artin

\artinpart{Hardware}

One $8 \times$ A100(40GB-memory) node.

CPU: At least 64 cores.

Memory: At least 200 GB.

Storage: Depending on the model and training epochs, recommend at least at least 350 GB for a 7B model and 700 GB for 14B model.

\artinpart{Software}

The used software are CUDA-12.8, DeepSpeed-v0.17.2, PyTorch-2.7.1, transformers-4.55.0, pydantic-2.9.2, flash\_attn-2.6.3. The used models are Llama-3.2-1B, Llama-3.1-8B, and Qwen-2.5-7B.

(https://huggingface.co/meta-llama/Llama-3.2-1B), 

(https://huggingface.co/meta-llama/Llama-3.1-8B), 

(https://huggingface.co/Qwen/Qwen2.5-7B).

\artinpart{Datasets / Inputs}
We use two medical datasets: 

(i) PubMed-Summarization (https://huggingface.co/datasets/ccdv-/pubmed-summarization)

(ii) MedQA (https://huggingface.co/datasets/Malikeh1375/medical-question-answering-datasets)

\artinpart{Installation and Deployment}
First, using anaconda to create an environment using python version 3.11. Then using git to clone our artifacts and using 'pip install -r requirements.txt' to install our dependencies.

The examples are in the folder '/example'. Then, modifying the YAML file to whatever you like.
Next, modifying the configuration in the top of this start\_merge.py file. (e.g. CHECKPOINT\_PATH). And finally, run these steps in example.ipynb. 

For the benchmark running tasks, we use the open source project called lm-evaluation-harness (https://github.com/EleutherAI/lm-evaluation-harness). Please follow the instructions of this project to install and use.

\artcomp

The workflow consists of 3 tasks:
\begin{enumerate}
    \item \textbf{$T_1$.} Run an LLM training job that generates multiple checkpoints. Our package can be used as a submodule within a partial checkpointing framework, and is fully compatible with checkpoints produced in this form.  

    \item \textbf{$T_2$.} To configure and start our program, the user should provide: the path to the previous checkpoints, the output path for the new "Frankenstein" checkpoint, the number of GPUs, the failure step number, and the number of hidden layers in the training model.  
    If partial checkpoints are used, the user may also provide the JSON file generated by the partial checkpointing system. In this case, our tool will automatically generate a corresponding YAML file. Otherwise, the user can manually write a YAML file specifying the layers to be merged. Finally, LLMTailor will select layers from different checkpoints and assemble them into a complete new checkpoint.  
    
    \item \textbf{$T_3$.} Use the path of the newly generated checkpoint to resume training. Observe whether the loss curves align with those of uninterrupted training, thereby confirming recovery from failure. For the final stage, evaluate the model performance on the benchmarks mentioned in the paper to validate correctness and consistency.  
\end{enumerate}

\artout

Running Task $T_1$ will generate multiple checkpoints together with an optional JSON file that records the partial checkpointing decisions. Different layers are checkpointed at different steps, and the detailed information is logged in the JSON files. By comparing the size of these checkpoints, one can verify that when only a subset of layers is saved, the resulting checkpoint is smaller than a full checkpoint.

Task $T_2$ produces a complete checkpoint folder. Users can confirm correctness by comparing its size and file structure against a standard full checkpoint generated without selective saving.

Executing Task $T_3$ demonstrates recovery from a simulated failure. Training should resume exactly at the failure step, and the reported loss and downstream performance metrics will confirm that recovery is almost consistent with uninterrupted training.

\end{document}